\documentclass[showpacs,preprint,amsmath,amssymb,epsf]{revtex4}
\usepackage{latexsym}
\usepackage{graphics}
\usepackage{epsfig}
\usepackage{dcolumn}
\def\br{\begin{eqnarray}}
\def\er{\end{eqnarray}}
\def\be{\begin{equation}}
\def\ee{\end{equation}}

\def\G{\Gamma}

\def\m{\mu}

\def\({\left(}
\def\){\right)}
\def\<{\left\langle}
\def\>{\right\rangle}

\def\S{\Sigma}

\begin{document}
%
%
\draft  
\title{The Energy Criterion and Dynamical Symmetry Breaking in a $SU(3)_{{}_{L}}\otimes U(1)_{{}_{X}}$ Extension of the Standard Model }
\author{A. Doff$^{a}$} 
\affiliation{$^a${\small Universidade  Tecnol\'ogica  Federal do Paran\'a - UTFPR - COMAT, Via do Conhecimento Km 01, 85503-390, Pato  Branco, PR, Brazil}}
\email{agomes@utfpr.edu.br}

\date{\today}
\begin{abstract} 
The coupling constants $g_{{}_{L}}$ and $g_{{}_{X}}$  of some versions of  the $SU(3)_{{}_{L}}\otimes U(1)_{{}_{X}}$ extension of the standard model are related through to relationship $g^2_{X}/g^2_{{L}}= \sin^2\theta_{{}_{W}}/(1 - 4\sin^2\theta_{{}_{W}})$. This fact suggest that  the $SU(3)_{{}_{L}}\otimes U(1)_{{}_{X}}$  gauge symmetry  in this class of models can be broken dynamically to the standard model at TeV scale without requiring the introduction of fundamental scalars. This possibility was investigated by Das and  Jain  who considered only the first version of this class of models. In this brief report  we discuss  an energy criterion to verify  the most probable version of the  $SU(3)_{{}_{L}}\otimes U(1)_{{}_{X}}$  model that is realized in nature.
\end{abstract}
\maketitle

\par The Standard Model of elementary particles is in excellent agreement with the experimental data and has explained many features of particle physics along the years. Despite the success there are some points in the model as, for instance,  the flavor problem  or the enormous range of masses  between the lightest and heaviest  fermions and other peculiarities the could be better explained with the introduction of new fields and symetries. One of the possibilities in this direction is to assume an extension of the Standard Model based on $G_{3n1} \equiv  SU(3)_{{}_{C}}\otimes SU(n)_{{}_{L}}\otimes U(1)_{{}_{X}}$\cite{felice1, tonasse, felice2}, where $n=3,4$. This class of the models predicts interesting new physics at TeV scale\cite{trecentes} and address some fundamental questions that does not can be explained in the framework of the Standard Model. As a brief example we can mention the flavor problem\cite{dp1} and the question of electric charge quantization\cite{dp2}.

\par One interesting feature of some versions of these models\cite{felice1, tonasse} is the following relationship among the coupling constants $g_{{}_{L}}$ and $g_{{}_{X}}$ associated to the gauge group $SU(3)_{{}_{L}}\otimes U(1)_{{}_{X}}$ 
\begin{equation}
\frac{\alpha_{X}}{\alpha_{{L}}} = \frac{\sin^2(\theta_{{}_{W}})}{1 - 4\sin^2(\theta_{{}_{W}})}
\label{eq1}
\end{equation}
\noindent where $\alpha_{i} = g^2_{i}/4\pi$, with  $i= X, L$ and  $\sin(\theta_{{}_{W}})$ is the electroweak mixing angle. Then, in a high energy scale, when $\sin^2(\theta_{{}_{W}})(\mu)\approx 1/4$, the coupling constant $g^2_{{}_{X}}$ becomes very strong. The energy scale where the theory  becomes non-perturbative may be estimated as being  of order  few TeVs, and this fact suggest  that  the gauge symmetry $SU(3)_{{}_{C}}\otimes SU(3)_{{}_{L}}\otimes U(1)_{{}_{X}}$ of this class of models maybe  be broken  to $SU(3)_{{}_{C}}\otimes SU(2)_{{}_{L}}\otimes U(1)_{{}_{Y}}$ without requiring the introduction of fundamental scalars.

\par In Ref\cite{Das} the  authors  investigated this possibility, however, in that work only the first version of these models was considered \cite{felice1}. As we emphasize in the text, there are  other versions  of 3-3-1 models that have  the relationship show in Eq.(1) among the coupling constants $\alpha_{X}$ and $\alpha_{{L}}$\cite{tonasse}. In this work  we discuss an energy criterion\cite{dn} to  select the most probable version of the  3-3-1 model. We show  that just  only version\cite{tonasse} of these models lead to a deeper minimum of the effective potential. Therefore, only in this version the coupling  constant $\alpha_{X}$ would  be strong enough, and energetically preferred, in order to promote 
the dynamical symmetry breaking.


\par We will begin  writing  the Schwinger-Dyson equation  for quarks considering only  the $U(1)_{X}$ interaction  once this is the dominant contribution
\begin{equation}
S^{-1}(p) = \slash{\!\!\!p} -i\int\frac{d^4q}{(2\pi)^4}\Gamma_{\mu}(p,q)S(q)\Gamma_{\nu}D^{\mu\nu}_{{}_{M_{Z'}}}(p-q)
\label{sde}
\end{equation}
\noindent where  we assumed the rainbow approximation for the vertex $\Gamma_{\mu, \nu}$, with 
$\Gamma_{\mu,\nu} = g_{{}_{V}}\gamma_{\mu,\nu} + g_{{}_{A}}\gamma_{\mu,\nu}\gamma_{5}$, $g_{{}_{V}} = g^2_{X}(X_{{}_{L}} + X_{{}_{R}})/2$ and $g_{{}_{A}} = g^2_{X}(X_{{}_{R}} - X_{{}_{L}})/2$. $X_{L}$ and $X_{R}$ are respectively  the $U(1)_{{}_{X}}$ charges attributed to the  chiral components of the exotic quarks $J_{1L}$ and $J_{1R}$. 
\par With the purpose of simplifying the calculations is convenient choose the Landau gauge. In this case the $Z'$ propagator  can be written in the following form 
$$
iD^{\mu\nu}_{{}_{M_{Z'}}}(p - q) = -i\frac{\left[g_{\mu\nu} - (p-q)_{\mu}(p - q)_{\nu}/(p-q)^2\right]}{(p - q)^2 - M^2_{Z'}}.
$$
\noindent Writing the  quark propagator  as $i{S}^{\,-1}(p) = i(\not{\!p} - \Sigma(p^2))$, and considering the equation above, we  can write  
\begin{eqnarray}
\Sigma(p^2) =  - ia\int d^4q \frac{\Sigma(q^2)}{[q^2 - \Sigma^2(q^2)]}\frac{1}{[(p - q)^2 - M^2_{Z'}]}
\label{eq5}
\end{eqnarray}
\noindent where $a = \frac{3g^2_{{}_{X}}X_{{}_{L}}X_{{}_{R}}}{(2\pi)^4}$. 
\par The dynamical mass generated for the $Z'$ boson can be estimated as $M_{Z'} \sim \mu_{X}$, where $\mu_{X} \sim O(TeV)$ is the energy scale where the $U(1)_{X}$ interaction becomes  sufficiently strong to  break dynamically the quiral and gauge symmetries. The Eq.(3) is one nonlinear integral equation and can be reduced to a nonlinear differential equation in momentum space which can be solved only numerically.  However,  we can use the  following linearized version of this last one 
\begin{eqnarray} 
\frac{d}{dp^2}\left[(p^2 + \mu^2_{{}_{X}})^2\frac{d\Sigma(p^2)}{dp^2}\right] = -a\frac{p^2\Sigma(p^2)}{(p^2 + \mu^2_{{}_{X}})}\nonumber \\
\end{eqnarray}
\noindent  as a good approximation, where we substituted in the denominator  $\Sigma^2(p^2)$ by $\mu^2_{X}$. Once only the exotic quarks $J$ will acquire mass at this scale the dynamical mass generated for such particles will be of the same order that the mass generated for the new  gauge bosons $(V^{\pm}, Z')$, justifying  our  approximation. The most general solution for this equation can be written as 
\begin{equation}
\Sigma(p^2) = \frac{f(p^2)}{2n}\left[C_{n}J_{n}[f(p^2)]\Gamma(n) -  C_{m}J_{m}[f(p^2)]\Gamma(m) \right]
\label{SSD}
\end{equation}
\noindent where  $J_{n,m}[z]$ are  Bessel functions, $\Gamma(n,m)$ is the Gamma function and $ C_{n,m}$ are constants of integration. For convenience, we defined  
$$ 
f(p^2) = \left(\frac{4a\mu^2_{X}}{p^2 + \mu^2_{X}}\right)^{\frac{1}{2}}
$$
\noindent with  $n = -m \equiv \sqrt{1 - 4a} $. Eq.(\ref{SSD}) has two asymptotic solutions
\br 
&&\Sigma(p^2)_{1} \sim  \frac{\mu^3_{X}}{p^2}\left(\frac{p^2}{\mu^2_{X}}\right)^{a} \\ 
&&\Sigma(p^2)_{2} \sim \mu_{X}\left(\frac{p^2}{\mu_{X}}\right)^{-a}. 
\label{easympt}
\er 
\noindent which are named in the literature  respectively as Regular and Irregular  solutions\cite{natale,ln}. Considering  the running  of the  $U(1)_{{}_{X}}$ coupling constant 
\be 
\alpha_{{}_{X}}(p^2) = \frac{\alpha_{{}_{X}}(\mu^2)}{1 + \alpha_{{}_{X}}(\mu^2) b_{{}_{X}}\ln(\mu^2/p^2)}
\ee  
\noindent where $ b_{{}_{X}} \equiv  \Sigma X^2/{6\pi}$, the asymptotic solutions of the Eq.(6) and (7)  can be written as 
\br
&&\Sigma(p^2)_{1} \sim  \frac{\mu^3_{X}}{p^2}\left(\frac{\alpha_{{}_{X}}(p^2)}{\alpha_{{}_{X}}(\mu^2_{X})}\right)^{c} \\  
&&\Sigma(p^2)_{2}  \sim  \mu_{X}\left(\frac{\alpha_{{}_{X}}(p^2)}{\alpha_{{}_{X}}(\mu^2_{X})}\right)^{-c}. 
\label{easympt2}
\er 
\noindent where $c = \frac{9X_{L}X_{R}}{2\Sigma X^2}$ and in the expressions above $\Sigma X^2$  is the sum of the square of the $U(1)_{{}_{X}}$ charges of the models to be considered in this work. There is an restriction  about the Irregular solution,  Eq.(10). For this solution  it is necessary that  $c > 1/2$ \cite{ln}. If we consider the formal equivalence between the solution of the Schwinger-Dyson equation with the Bethe-Salpeter one for pseudo-scalar bound states, the above restriction indicates the condition for wave function normalization of the Goldstone bosons. In all the models that we will be  considered in this work we have  $c < 1/2$, and for this reason we will just consider the fermionic self-energy Eq.(9). In the paragraph  below we will  compute the vacuum energy  for this fermionic self-energy making use of the effective potential for composite operators. 




\par The effective potential for composite operators is given by the following expression\cite{cornwall1}
\be
V(S,D) = - \imath \int \frac{d^4p}{(2\pi)^4}Tr ( \ln S_0^{-1}S - S_0^{-1}S + 1)
 +\,\,V_2(S,D),
\label{efpot}
\ee 
where in this expression $S$ and $D$ are the complete propagators of fermions and gauge bosons and $S_0$, $D_0$, are  the corresponding bare propagators.  The function $V_2(S,D)$ is given by the two-particle irreducible vacuum diagram  depicted in the figure 1. 
\par  The expression  for  $V_2(S,D)$ can be represented   analytically in the Hartree-Fock approximation by  the following equation 
\be
\imath V_2(S,D) = - \frac{1}{2} Tr(\G S \G S D)
\ee 
\begin{figure}[t]
\begin{center}
\epsfig{file=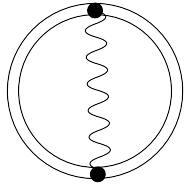,width=0.15\textwidth}
\caption{Diagramm  for  two loops $(V_{2})$ contribution to the effective potential}
\end{center}
\end{figure}
\noindent  where for simplicity in this equation  we have not written the gauge and Lorentz indices, as well as the momentum integrals and we are representing the fermion proper vertex by $\Gamma$. 
\par We want to determine numerically the vacuum expectation value for the fermionic self-energy  given by the Eq.(9), for the models described in the Ref.\cite{felice1}(A,B) and Ref.\cite{tonasse}(C). However, it is better to compute the vacuum energy density, which is given by the effective potential calculated at minimum subtracted by its perturbative part which does not contribute to dynamical mass generation\cite{cornwall1,castorina}
\be
\< \Omega \> = V_{min}(S,D) - V_{min}(S_p,D_p),
\ee 
\noindent where we indicate in the above expression the perturbative counterpart of $S$ and $D$ respectively by $S_p$, $D_p$. $V_{min}(S,D)$ is obtained substituting the SDE, Eq.(\ref{sde}), in  the Eq.(\ref{efpot})  and  we can verify that in the chiral limit $S_p = S_0$. The complete fermion propagator $S$ is related to the free propagator by the equation $S^{-1} = S_0^{-1} - \Sigma$, with $S_0 = \imath /\not \!\! p$ and  we chose to work in the  Landau gauge. After  a  Euclidean rotation,  we find that $\Omega_{min} \equiv \< \Omega \>$ is equal to\cite{castorina}
\be
\Omega_{min}  = -2 \int \frac{d^4p}{(2\pi)^4} \,V(p^2, \Sigma)
\ee
\noindent  where we have defined  the function $V(p^2, \Sigma)$ as 
$$ 
V(p^2, \Sigma) = \left[ \ln ( \frac{p^2 + \Sigma^2}{p^2} ) - \frac{\Sigma^2}{p^2 + \Sigma^2} \right].
$$
\par We can still expand  $\Omega_{min}$ in powers of $\Sigma^2/{p^2}$, so that
\be
\Omega_{min}  \approx  - \int \frac{d^4p}{(2\pi)^4} \frac{\S^4}{p^4}.
\ee
\par To obtain an analytical formula for the  vacuum energy density  we will  consider  the
substitution $x \rightarrow \frac{p^2}{\m^2_{X}}$ in the Eqs.(9) and (15),  and we will  assume the following   Mellin transform\cite{cs}
\be
\left[ 1 + \kappa \ln {x} \right]^{-\epsilon} =
\frac{1}{\Gamma ({\epsilon})}\int_0^\infty d\sigma \, e^{-\sigma}
\left( {x} \right)^{-\sigma \kappa} \sigma^{\epsilon - 1}
\label{mt}
\ee 
\noindent  that  will simplify considerably the calculation.  In this Mellin transform  we identified $\kappa =  -\alpha_{{}_{X}}b_{{}_{X}}$ and $\epsilon = 4c$. Then, after we substitute Eq.(9) in to Eq.(15), and perform the integration we obtain
\br
\Omega_{min} \approx  - \frac{\mu^4_{X}\zeta}{32\pi^2}\left[1 + \frac{1}{8\pi^2\zeta}\frac{\Sigma X^2}{X_{{}_{L}}X_{{}_{R}}} + O(\frac{1}{\zeta(\Sigma X^2)^2})...\right].\nonumber \\
\er 
\noindent Where $\zeta \equiv \left(1 + \frac{3}{4\pi}\right)$, and to obtain this last equation we made use of the  scaling law $\frac{g^2_{X}X_{{}_{L}}X_{{}_{R}}}{4\pi}\approx 1$ \cite{susskind}.
\par In table I we present the  weak hypercharge content attributed  to each model and  the respective value obtained for the minimum of the potential. As it is possible to verify  the deepest minimum of energy happens  only for the model C because at scale $\mu_{{}_{X}}$ this model take to $U(1)_{X}$  coupling  much more stronger and close to the critical value $\alpha_{c}$   necessary to promote the dynamical symmetry breaking.

\begin{table}[b]
\begin{ruledtabular}
\begin{tabular}{ccc}
$32\pi^2\Omega_{min}/\mu^4_{{}_{X}}$ & Model & $U(1)_{X}$ Charges \\ \\
\hline
               &          & leptons:           \\ 
               &          & $X_{l_{aL}} =  0$           \\
               &          & quarks:         \\
    -1.535     &   A,B    & $X_{Q_{1L}} = \,\,\,\, 2/3$\,\,\,\,,\,\,$X_{u_{1R}} = 2/3 $ \\
               &          & $X_{d_{1R}} = -1/3$\,\,\,,\,\,\,$X_{J_{1R}} = 5/3 $ \\ 
               &          & $X_{Q_{iL}} = -1/3$\,\,\,,\,\,\,$X_{u_{iR}} = 2/3 $ \\
               &          & \,\,$X_{d_{iR}} = -1/3$\,\,\,,\,\,\,$X_{J_{iR}} =-4/3 $  \\ \hline                
               &          & leptons:           \\ 
               &          & $X_{l_{aL}} = 0$\,\,,\,\,$X_{l_{aR}}= -1$ \\
               &          & $X_{E_{aR}} = 1 $  \\
               &          & quarks:         \\
    -1.605     &     C    & $X_{Q_{1L}} = \,\,\,\, 2/3$\,\,\,\,,\,\,$X_{u_{1R}} = 2/3 $ \\
               &          & $X_{d_{1R}} = -1/3$\,\,\,,\,\,\,$X_{J_{1R}} = 5/3 $ \\ 
               &          & $X_{Q_{iL}} = -1/3$\,\,\,,\,\,\,$X_{u_{iR}} = 2/3 $ \\
               &          & \,\,\,$X_{d_{iR}} = -1/3$\,\,\,,\,\,\,$X_{J_{iR}} =-4/3 $  \\            
\end{tabular}
\end{ruledtabular}
\caption{In the above table $i = 2,3$ labels the second and third quark families and $a = 1..3$. $X_{l_{aL}}$, $X_{Q_{1L}}$ and  $X_{Q_{iL}}$ represent, respectively, the hipercharges attributed to ${\bf 3}$ and ${\bf 3^*}$ of the leptons$(l)$ e quarks$(Q)$.  We also show the hipercharge content  attributed to the models A,B and C.  The fermionic content associated to the model B is the same as in model A, however, it is the third quark generation that will transform as ${\bf 3^{*}}$, the first and second generation will transform as ${\bf 3}$.}
\end{table}  
  
\par As the authors  of Ref.\cite{Das} argued, the  gauge symmetry breaking  of  $SU(3)_{{}_{L}}\otimes U(1)_{{}_{X}}$ in 3-3-1 models  can be implemented dynamically  because the scale of a few TeVs, $\mu_{X}$,  the $U(1)_{X}$ coupling constant becomes strong as we approach the peak existent in  Eq.(1). The exotic quarks $J$ introduced in these models  will form a condensate $\langle\bar{J}J\rangle$ breaking $SU(3)_{{}_{L}}\otimes U(1)_{{}_{X}}$  to  $SU(2)_{{}_{L}}\otimes U(1)_{{}_{X}}$ at this scale.
\par The electroweak symmetry could be broken dynamically by a top condensate\cite{hill}. In this case, as Das and Jain argued, it would be necessary to introduce  new  exotic quarks $\chi_{L}$ and $\chi_{R}$  in the model in order to maintain the top quark mass around $170GeV$, which is an interesting possibility that we intend to explore in the future.
\par  In this work we show  that just  one version\cite{tonasse} of this class of models lead to a deeper minimum of the effective potential, establishing a criterion for the choice  of the most probable version of the  $SU(3)_{{}_{L}}\otimes U(1)_{{}_{X}}$  model that is realized in nature. 
\par After the integration  of equation (\ref{vminf2}), we obtain the value for the vacuum energy density(minimum of energy), $\Omega_{min}$, for models of the type $A$ to $C$, as shown  in table I.  At the scale $\mu_{X}$, the coupling constant $\alpha_{X}$ of the $U(1)_{{}_{X}}$ group becomes strong enough to promote  the dynamical symmetry breaking  of the model. However, this happens only for the version C, which is the one that corresponds to the deepest state of energy.

\begin{acknowledgments}
I would like to thank A. A. Natale for discussions and the Funda\c c\~ao de Apoio ao Desenvolvimento Cient\'{\i}fico e Tecnol\'ogico do Paran\'a (Funda\c c\~ao Arauc\'aria) by  financial support. 
\end{acknowledgments}

\pagebreak

\end{document}